# A CO(3–2) Survey of
# Nearby Mira Variables

*K. Young*

Caltech Submillimeter Observatory, P.O. Box 4339, Hilo, HI 96720

1995–2

## ABSTRACT

A survey of CO(3–2) emission from optically visible oxygen–rich Mira variable stars within 500 pc of the sun was conducted. A molecular envelope was detected surrounding 36 of the 66 stars examined. Some of these stars have lower outflow velocities than any Miras previously detected in CO. The average terminal velocity of the ejected material was $7.0 \ \mathrm{km \, s^{-1}}$, about half the value found in Miras selected by infrared criteria. None of the stars with spectral types earlier than M 5.5 were detected. The terminal velocity increases as the temperature of the stellar photosphere decreases, as would be expected for a radiation driven wind. Mass loss rates for the detected objects were calculated, and it was found that there is no correlation between the infrared color of a Mira variable, and its mass loss rate. The mass loss rate is correlated with the far infrared luminosity, although a few stars appear to have extensive dust envelopes without any detectable molecular wind. A power–law relationship is found to hold between the mass loss rate and the terminal velocity of the ejected material. This relationship indicates that the dust envelope should be optically thick in the near infrared and visible regions of the spectrum when the outflow velocity is $\gtrsim 17 \ \mathrm{km \, s^{-1}}$. At the low end of the range of outflow velocities seen, the dust drift velocity may be high enough to lead to the destruction of the grains via sputtering. Half of the stars which were detected were re–observed in the CO(4–3) transition. A comparison of the outflow velocities obtained from these observations with those obtained by other investigators at lower frequencies shows no evidence for gradual acceleration of the outer molecular envelope.

## 1. Introduction

Over the last 25 years the new infrared surveys, particularly the Two–Micron Sky Survey (Neugebauer & Leighton 1969, hereafter TMSS), the Air Force Geophysical Laboratory survey (Price & Walker 1975, hereafter AFGL), and the *IRAS* Point Source Catalog (Neugebauer *et al.* 1984, hereafter PSC), have revealed the presence of tens of thousands of cool stars surrounded by dusty envelopes. After each of these surveys was completed, observers examined the most prominent of the newly detected objects in the radio regime, and were frequently able to detect molecular line emission from molecules such as OH, HCN and CO. From this confluence of infrared and radio data arose a compelling model for mass loss in highly evolved stars. According to this model, dust condenses in the outer envelope of a cool red giant star. This dust is then accelerated by radiation pressure to a velocity of $\sim 15\ \mathrm{km\,s^{-1}}$. The dust in turn accelerates gaseous material by colliding with it (see a review by Olofsson 1988).

This paper presents the results of a survey of Mira variables with distances of 500 pc or less, which were selected without regard to their infrared characteristics. There are several reasons for carrying out such a survey. First, it is possible that the radiation pressure mechanism described above is not the only one driving mass loss from red giant stars. Mechanisms which do not require the formation of dust, such as radial pulsations (Wood 1979), momentum transfer from Alfvén waves (Hartmann & MacGregor 1980) or from sound waves (Fusi–Pecci & Renzini 1975, Pijpers & Hearn 1989) have been proposed for driving AGB winds. Finding a large, expanding molecular envelope surrounding a highly evolved star without a massive dust envelope would be strong evidence for an alternative mass loss mechanism. Secondly, because previous surveys tended to concentrate on the very strongest infrared sources, stars with small, but detectable, mass loss rates may be under–represented in their compilations of detections. The final reason for surveying nearby Miras is that they obey a period – luminosity relationship. This allows their distances to be estimated more accurately than many of the other types of stars which have been detected in the surveys of bright infrared objects. The derived values of many important physical parameters, including the absolute luminosity and the mass loss rate, are sensitive to errors in the estimated distance.

Examining the CO(3–2) line, rather than a lower J transition, offers several advantages. The first is that higher J transitions have higher critical densities for collisional excitation. For this reason the emission from molecular clouds is less extensive, and the spectra of Miras near the Galactic plane are less apt to be contaminated by galactic emission. Secondly, the antenna temperature of the 3–2 line will in general be higher than in the lower J lines (Morris 1980). While the rotational transitions above 3–2 may have even higher brightness temperatures, particularly for optically thin envelopes, the earth's atmosphere is much less transparent where these transitions are found, even on high mountain sites. The 3–2 line may therefore allow the most sensitive ground–based searches for cool circumstellar envelopes. Finally, for those objects which have already been detected in lower J surveys, the availability of spectra covering several CO transitions allows the models of molecular excitation to be better constrained, and variations of the CO envelope at different radii can be studied (van der Veen & Olofsson 1989).

## 2. Observations

The candidate objects for this survey where selected from the *Revised Catalog of Spectra of Mira Variables of Types Me and Se* (hereafter RCMV) (Keenan *et al.* 1974). The spectral types and periods of the stars were used to calculate an absolute visual magnitude, using the formulae derived by Celis (1980). This calculated absolute magnitude was used with the apparent magnitude, to calculate the star's distance. The star's sky position and distance were then used to calculate the amount of absorption, using Parenago's formula with coefficients from Sharov (1964). Finally the amount of absorption was used to derive a new distance estimate. This process of estimating the distance was iterated until it converged. Distances to a smaller set of stars were calculated in this way by Celis (1981), who estimated the mean error in derived distance values to be 19 %. However distances to many of these stars were later calculated by Jura and Kleinmann (1992) using infrared data, and their distance estimates differ from those of Celis by 50% on average. 79 of the stars in the RCMV were found to have a distance of 500 pc or less. 66 of these stars were observed.

All observations presented here were made at the Caltech Submillimeter Observatory,* (CSO) on Mauna Kea, Hawaii. Data were taken during January, May and August of 1990, and during February through August of 1991. The $^{12}C^{16}O$(3–2) line at 345.796 GHz was observed with an SIS receiver (Ellison & Miller 1987) having a double sideband temperature of about 200 K. A 1024 channel AOS spectrometer was used, which has a total bandwidth of 500 MHz and a velocity resolution of 0.9 km s$^{-1}$. In most cases, scans were taken while position switching at 0.05 Hz between the source position and 2 off–source positions located 3 arc minutes away in azimuth. For weak objects at low galactic latitudes, the scans used 4 off-source positions spaced symmetrically about the source, and 30 arc seconds away (1.5 beam widths), in order to remove any contamination from extended galactic emission. For those stars which appear in the SAO catalog, SAO coordinates and proper motions were used. For the remainder of the stars, coordinates were obtained from the *IRAS* PSC. Table 1 lists the names, coordinates and on–source integration times for the stars. For the stars which had not previously been detected in a rotational transition of CO, the radial velocity used for receiver tuning was obtained from the RCMV. If the star showed only weak CO emission (T$_{mb}$ ≤ 200 mK), it was re–observed on at least one additional night in order to confirm the detection. Two of the stars which were not detected, R Car and R Cen, never rise above 10 degrees from the horizon when observed at the CSO. Since telescope pointing can be problematic at very low elevations, the pointing was checked after these two stars were observed by observing the strong source IRC+10°216 as it passed through an elevation of 10 degrees. Both the CO(3–2) and CS(7–6) transitions were seen, indicating that errors in pointing were not responsible for these non–detections.

In April – May of 1993, and January 1994, nineteen of the stars with strong CO(3–2) emission were examined again with a second SIS receiver (Walker *et al.* 1992), tuned to the CO(4–3) line at 461.041 GHz, where it had a double sideband temperature of ∼ 250 K. The system temperature was ∼ 2500 K, allowing lines as weak as 0.1 K to be detected in an hour of integration. Because the LSR velocity of each of the stars was known from the CO(3–2) profile, a narrower AOS spectrometer having a total bandwidth of 50 MHz, was used for most of these observations, providing much greater velocity resolution. The 500 MHz spectrometer was used for R Cas, because its emission covered too large a velocity range to fit comfortably into the 50 MHz backend. All but one of the nineteen stars were detected in this higher transition.

### 3. Calibration

The receiver was sensitive to radiation in both sidebands, which were separated by 2.8 GHz. Indeed, when CO(3–2) was located in the upper sideband, $^{29}SiO$(8–7) was detected in the lower sideband for several objects ( R Ser, R Leo and especially W Hya). The relative sensitivity of the receiver to the two sidebands is difficult to determine, and it changes when the receiver's mechanical tuning structures are moved. For this reason, the carbon star IRC+10°216 was observed whenever the receiver was mechanically retuned, to assess its sensitivity to the signal sideband. Corrected for the main beam efficiency as measured

* The CSO is operated by the California Institute of Technology under funding from the National Science Foundation, Contract #AST–93–13929.

## Table 1. – Survey Star Positions, Integration Times and Spectrum Noise Levels

Nondetections are listed in parenthesis

| Star | R.A. (1950) | Decl. (1950) | Time (min.) | rms (mK) | Star | R.A. (1950) | Decl. (1950) | Time (min.) | rms (mK) |
|---|---|---|---|---|---|---|---|---|---|
| (S Scl) | 00 12 51.11 | −32 19 22.5 | 27.2 | 89 | W Hya CO(4–3) | | | 16.9 | 410 |
| (T And) | 00 19 46.43 | 26 43 10.3 | 7.1 | 180 | (R CVn) [A,D] | 13 46 48.10 | 39 47 28.0 | 51.8 | 31 |
| T Cas | 00 20 31.50 | 55 30 56.0 | 23.3 | 100 | (R Cen) | 14 12 56.90 | −59 40 55.2 | 34.9 | 170 |
| CO(4–3) | | | 54.4 | 210 | (U UMi) [A] | 14 16 13.40 | 67 01 29.0 | 16.8 | 46 |
| (TU And) [A] | 00 29 44.70 | 25 45 12.0 | 13.6 | 98 | (S Boo) | 14 21 12.40 | 54 02 12.4 | 14.2 | 71 |
| W And | 02 14 22.70 | 44 04 26.0 | 8.4 | 75 | (R Boo) | 14 34 59.27 | 26 57 09.4 | 22.0 | 73 |
| o Cet | 02 16 49.04 | −03 12 13.4 | 5.2 | 210 | (Y Lib) [A] | 15 09 02.50 | −05 49 27.0 | 11.7 | 76 |
| CO(4–3) | | | 9.7 | 380 | S CrB [C] | 15 19 21.53 | 31 32 46.5 | 9.1 | 82 |
| T Ari [C] | 02 45 32.04 | 17 18 07.2 | 108.7 | 18 | CO(4–3) | | | 63.4 | 100 |
| R Hor | 02 52 13.00 | −50 05 32.0 | 3.9 | 300 | RS Lib [A,C] | 15 21 24.30 | −22 44 06.0 | 24.6 | 50 |
| U Ari [A,C] | 03 08 16.20 | 14 36 40.0 | 31.7 | 46 | CO(4–3) | | | 25.9 | 118 |
| S Pic [A] | 05 09 37.20 | −48 34 00.0 | 50.5 | 59 | (S UMi) [A] | 15 31 24.10 | 78 47 54.0 | 19.4 | 100 |
| R Aur | 05 13 15.26 | 53 31 56.6 | 8.7 | 110 | (T Nor) | 15 40 11.71 | −54 49 44.8 | 31.7 | 160 |
| CO(4–3) | | | 31.1 | 140 | R Ser [C] | 15 48 23.24 | 15 17 02.7 | 24.6 | 36 |
| (W Aur) [A,B,C] | 05 23 31.20 | 36 51 40.0 | 28.5 | 47 | (CO(4–3)) | | | 64.7 | 85 |
| S Ori [B,C] | 05 26 32.68 | −04 43 51.7 | 14.2 | 51 | (Z Sco) [A] | 16 03 04.10 | −21 35 55.0 | 6.5 | 86 |
| CO(4–3) | | | 69.2 | 170 | RU Her [A,C] | 16 08 08.60 | 25 11 59.0 | 14.2 | 79 |
| U Aur | 05 38 53.50 | 32 00 57.0 | 12.9 | 98 | CO(4–3) | | | 64.7 | 128 |
| CO(4–3) | | | 47.9 | 67 | U Her [C] | 16 23 34.86 | 19 00 18.0 | 12.9 | 68 |
| U Ori | 05 52 51.00 | 20 10 06.2 | 16.8 | 94 | CO(4–3) | | | 24.6 | 110 |
| CO(4–3) | | | 60.8 | 110 | (S Her) | 16 49 37.14 | 15 01 27.8 | 19.4 | 68 |
| (V Mon) | 06 20 11.69 | −02 10 07.0 | 19.4 | 44 | RS Sco [B,C] | 16 51 59.82 | −45 01 22.9 | 29.1 | 47 |
| (X Gem) [A,B] | 06 43 55.20 | 30 19 53.0 | 40.1 | 25 | RR Sco [B,C] | 16 53 26.31 | −30 30 08.3 | 31.1 | 33 |
| (Y Mon) [A] | 06 54 05.80 | 11 18 33.0 | 50.5 | 97 | (R Oph) | 17 04 53.43 | −16 01 39.5 | 24.6 | 53 |
| (RS Mon) [A] | 07 04 49.30 | 05 03 58.0 | 50.5 | 150 | X Oph [B] | 18 35 57.51 | 08 47 19.7 | 8.4 | 86 |
| S CMi [A] | 07 30 00.00 | 08 25 35.0 | 18.1 | 35 | R Aql | 19 03 57.70 | 08 09 08.0 | 3.2 | 120 |
| R Cnc [C] | 08 13 48.54 | 11 52 52.6 | 14.9 | 53 | CO(4–3) | | | 19.4 | 300 |
| W Cnc [A,C] | 09 06 57.90 | 25 27 06.0 | 78.9 | 24 | RT Aql [A,B,C] | 19 35 39.70 | 11 36 24.0 | 19.4 | 46 |
| (R Car) | 09 30 59.21 | −62 34 01.1 | 29.1 | 280 | χ Cyg | 19 48 38.30 | 32 47 08.0 | 6.5 | 190 |
| X Hya [C] | 09 33 06.92 | −14 28 03.5 | 45.3 | 28 | CO(4–3) | | | 19.4 | 240 |
| R LMi | 09 42 34.40 | 34 44 35.0 | 10.4 | 62 | (RR Sgr) [A] | 19 52 50.20 | −29 19 21.0 | 9.1 | 73 |
| CO(4–3) | | | 63.4 | 91 | Z Cyg [A,C] | 20 00 02.60 | 49 54 07.0 | 23.3 | 35 |
| R Leo | 09 44 52.23 | 11 39 41.9 | 16.9 | 50 | T Cep [A] | 21 08 53.30 | 68 17 12.0 | 12.9 | 110 |
| CO(4–3) | | | 43.2 | 110 | (S Lac) | 22 26 49.37 | 40 03 34.3 | 13.6 | 140 |
| (R UMa) | 10 41 07.67 | 69 02 22.3 | 62.2 | 30 | R Peg | 23 04 08.23 | 10 16 21.5 | 27.2 | 51 |
| (X Cen) [A] | 11 46 41.10 | −41 28 47.0 | 34.9 | 47 | W Peg [A] | 23 17 22.70 | 26 00 18.0 | 9.7 | 80 |
| (R Crv) | 12 17 02.32 | −18 58 41.7 | 19.4 | 79 | (S Peg) | 23 18 00.95 | 08 38 42.1 | 13.6 | 93 |
| (R Vir) | 12 35 57.67 | 07 15 47.0 | 16.8 | 89 | (R Aqr) | 23 41 14.18 | −15 33 41.8 | 72.5 | 36 |
| R Hya | 13 26 58.40 | −23 01 27.0 | 10.4 | 98 | R Cas | 23 55 51.68 | 51 06 36.4 | 5.2 | 100 |
| CO(4–3) | | | 20.7 | 270 | CO(4–3) | | | 15.5 | 400 |
| S Vir [A,C] | 13 30 22.80 | −06 56 18.0 | 23.9 | 47 | (Z Peg) [A] | 23 57 33.20 | 25 36 29.0 | 18.8 | 80 |
| CO(4–3) | | | 32.4 | 84 | (W Cet) | 23 59 33.64 | −14 57 15.2 | 45.3 | 33 |
| W Hya | 13 46 12.50 | −28 07 11.0 | 5.8 | 160 | | | | | |

none

[A]Coordinates for these objects are from the *IRAS* PSC.

[B]Four off–source positions spaced symmetrically 30 arc seconds away from the source position were used when these objects objects were observed, in order to minimize contamination from extended galactic emission, if any.

[C]These objects showed only weak CO emission, so they were observed and the detection was confirmed on at least 1 additional night.

[D]R CVn was marginally detected when initially observed, but was not detected when re–observed for confirmation.

on planets, the average value of $T_{mb}$ for IRC+10°216 for all tunings was 29 K. This value is in reasonably good agreement with the main beam temperature of 32 K, obtained by Wang *et al.* (1994), who employed a single–sideband filter in their measurements, which therefore did not suffer from any calibration errors introduced by unequal sideband gains. The value of $T_{mb}$ for IRC+10°216 is apt to be constant even if the central star's luminosity varies, since the rotational levels of CO in the envelope are populated primarily through collisions (Kwan & Hill 1977). For this reason, the temperature scales for each observation were scaled by the factor required to yield a $T_{mb}$ of 29.0 K for IRC+10°216. For the CO(4–3) observations, observations of IRC+10°216 were not available for each tuning. For these observations the antenna temperature was corrected only for the main beam coupling efficiency, which was measured to be 60% on both Jupiter and Mars.

## 4. Analysis

Linear baselines were fitted to regions of the spectra judged to be free of emission, and were then subtracted from the spectra. For those objects which were detected, a model line profile was fitted to the data. The model (Morris 1984) assumes that CO is expelled from the central star in a spherical shell, at a nearly constant velocity (which is much larger than either the thermal or turbulent velocities). The line's temperature profile T(v) has the form

$$T(v) = T_{mb}W \left( \frac{\exp(-\beta^2 W)}{\exp(-\beta^2)} \right) \left( \frac{1 - \exp(-\alpha/W)}{1 - \exp(-\alpha)} \right) \qquad \text{where} \quad W = 1 - \left( \frac{v - V_{LSR}}{V_o} \right)^2. \quad (1)$$

In the above equations $T_{mb}$ is the line's temperature at the systemic velocity $V_{LSR}$, and $V_o$ is the outflow velocity of the CO. $\alpha$ is related to the optical depth of the CO transition; large values correspond to high optical depths and give rise to parabolic line–shapes. Small values indicate that the transition is optically thin, and give rise to flat–topped lines. $\beta$ is the ratio of the size of the source to that of the telescope's beam. Large values of $\beta$ give rise to profiles with "horns" at $V_{LSR} \pm V_o$ on optically thin lines, or somewhat flattened parabolae in the optically thick case. Because none of the spectra show very pronounced horns, and most of them are too noisy to allow discrimination between a flattened parabola and a perfect one, for the purposes of profile fitting all sources are assumed to be completely unresolved ($\beta = 0$). The remaining 4 parameters were adjusted for their least–squares best fit. Table 2 lists which stars were detected, the results of the profile analysis, and the 95% confidence limits on the derived

**Table 2. – Line Fit Results**

| Star | $T_{mb}$ (K) | Area K km s$^{-1}$ | $(\alpha)$ | $V_{LSR}$ (km s$^{-1}$) | $V_o$ (km s$^{-1}$) |
|---|---|---|---|---|---|
| T Cas | $1.0 \pm 0.06$ | 17 | 0.8 | $-9.3 \pm 0.2$ | $10.5 \pm 0.5$ |
| (4–3) | $0.45 \pm 0.07$ | 5.5 | > 4 | $-5.7 \pm 0.6$ | $10.5 \pm 1.1$ |
| W And | $0.59 \pm 0.06$ | 7.0 | > 4 | $-36.7 \pm 0.3$ | $8.9 \pm 0.5$ |
| o Cet | $12 \pm 0.2$ | 75 | > 4 | $46.5 \pm 0.0$ | $4.8 \pm 0.0$ |
| (4–3) $^A$ | $10 \pm 0.15$ | 67.9 | 1.6 | $46.4 \pm 0.0$ | $4.7 \pm 0.1$ |
| T Ari | $0.10 \pm 0.01$ | 0.48 | 1.4 | $-0.8 \pm 0.3$ | $2.7 \pm 0.7$ |
| R Hor | $3.1 \pm 0.2$ | 23 | 1.9 | $35.7 \pm 0.2$ | $5.2 \pm 0.3$ |
| U Ari | $0.29 \pm 0.04$ | 1.7 | 1.3 | $-55.3 \pm 0.2$ | $4.0 \pm 0.3$ |
| S Pic | $0.33 \pm 0.04$ | 5.2 | 2.5 | $-0.9 \pm 0.5$ | $11.6 \pm 1.0$ |
| R Aur | $1.2 \pm 0.07$ | 16 | > 4 | $-3.4 \pm 0.2$ | $9.6 \pm 0.5$ |
| (4–3) | $0.90 \pm 0.04$ | 11.4 | > 4 | $-1.0 \pm 0.2$ | $10.0 \pm 0.3$ |
| S Ori | $0.45 \pm 0.04$ | 3.4 | > 4 | $14.5 \pm 0.2$ | $5.7 \pm 0.5$ |
| (4–3) | $0.39 \pm 0.07$ | 2.9 | 2.9 | $14.1 \pm 0.5$ | $6.4 \pm 0.9$ |
| U Aur | $0.41 \pm 0.04$ | 7.1 | 0.1 | $7.0 \pm 0.5$ | $9.2 \pm 0.5$ |
| (4–3) | $0.17 \pm 0.02$ | 2.2 | 3.8 | $6.8 \pm 0.5$ | $9.8 \pm 0.8$ |
| U Ori | $1.1 \pm 0.07$ | 11 | > 4 | $-37.5 \pm 0.2$ | $7.5 \pm 0.3$ |
| (4–3) | $0.72 \pm 0.04$ | 5.8 | > 4 | $-37.9 \pm 0.2$ | $6.5 \pm 0.3$ |
| S CMi | $0.31 \pm 0.02$ | 2.1 | 1.7 | $51.7 \pm 0.2$ | $4.7 \pm 0.3$ |
| R Cnc | $0.44 \pm 0.04$ | 2.7 | 0.3 | $15.0 \pm 0.2$ | $3.5 \pm 0.3$ |
| W Cnc | $0.097 \pm 0.02$ | 0.73 | 0.1 | $36.8 \pm 0.2$ | $4.2 \pm 0.5$ |
| X Hya | $0.097 \pm 0.01$ | 0.85 | 1.1 | $26.2 \pm 0.5$ | $5.9 \pm 1.1$ |
| R LMi | $0.82 \pm 0.02$ | 9.5 | > 4 | $1.3 \pm 0.2$ | $8.7 \pm 0.3$ |
| (4–3) | $0.63 \pm 0.02$ | 7.1 | > 4 | $0.6 \pm 0.2$ | $8.5 \pm 0.2$ |
| R Leo | $3.7 \pm 0.02$ | 37 | > 4 | $-0.4 \pm 0.0$ | $7.4 \pm 0.0$ |
| (4–3) | $2.8 \pm 0.04$ | 28.1 | 3.5 | $-0.4 \pm 0.0$ | $7.4 \pm 0.1$ |
| R Hya | $2.5 \pm 0.06$ | 37 | 2.3 | $-10.1 \pm 0.2$ | $10.5 \pm 0.2$ |
| (4–3) | $4.6 \pm 0.11$ | 46.1 | > 4 | $-10.0 \pm 0.1$ | $8.2 \pm 0.1$ |
| S Vir | $0.42 \pm 0.04$ | 2.5 | 2.3 | $9.8 \pm 0.2$ | $4.4 \pm 0.5$ |
| (4–3) | $0.35 \pm 0.04$ | 2.3 | > 4 | $10.1 \pm 0.2$ | $5.7 \pm 0.4$ |
| W Hya | $2.1 \pm 0.07$ | 29 | 0.3 | $39.9 \pm 0.2$ | $8.0 \pm 0.2$ |
| (4–3) | $3.1 \pm 0.09$ | 40.3 | 1.0 | $41.0 \pm 0.1$ | $8.4 \pm 0.2$ |
| S CrB | $0.88 \pm 0.04$ | 9.1 | > 4 | $1.5 \pm 0.2$ | $7.7 \pm 0.3$ |
| (4–3) | $0.60 \pm 0.04$ | 4.9 | > 4 | $1.4 \pm 0.2$ | $6.3 \pm 0.3$ |
| RS Lib | $0.35 \pm 0.04$ | 2.4 | 1.7 | $7.8 \pm 0.3$ | $4.9 \pm 0.5$ |
| (4–3) | $0.45 \pm 0.06$ | 3.1 | > 4 | $7.6 \pm 0.3$ | $5.5 \pm 0.5$ |
| R Ser | $0.37 \pm 0.02$ | 2.5 | > 4 | $31.7 \pm 0.3$ | $5.1 \pm 0.5$ |
| (4–3) | | | *not detected* | | |
| RU Her | $0.79 \pm 0.04$ | 9.4 | > 4 | $-11.4 \pm 0.3$ | $8.9 \pm 0.3$ |
| (4–3) | $0.53 \pm 0.04$ | 5.7 | 1.1 | $-11.1 \pm 0.2$ | $7.2 \pm 0.4$ |
| U Her | $0.47 \pm 0.04$ | 7.3 | > 4 | $-13.7 \pm 0.5$ | $11.5 \pm 0.6$ |
| (4–3) | $0.36 \pm 0.02$ | 4.9 | 0.1 | $-13.7 \pm 0.1$ | $7.4 \pm 0.2$ |
| RS Sco | $0.35 \pm 0.04$ | 2.3 | > 4 | $6.5 \pm 0.3$ | $5.0 \pm 0.3$ |
| RR Sco | $0.19 \pm 0.02$ | 0.85 | > 4 | $-28.9 \pm 0.3$ | $3.1 \pm 0.5$ |
| X Oph | $0.33 \pm 0.06$ | 2.9 | 1.3 | $-55.0 \pm 0.5$ | $6.1 \pm 0.9$ |
| R Aql | $3.5 \pm 0.06$ | 45 | > 4 | $46.0 \pm 0.2$ | $9.5 \pm 0.2$ |
| (4–3) | $3.1 \pm 0.09$ | 36.9 | > 4 | $47.1 \pm 0.1$ | $9.2 \pm 0.2$ |

*(continued on next page)*

**Table 2.** *(continued)*

| Star | $T_{mb}$ (K) | Area K km s$^{-1}$ | $(\alpha)$ | $V_{LSR}$ (km s$^{-1}$) | $V_o$ (km s$^{-1}$) |
|---|---|---|---|---|---|
| RT Aql | $0.22 \pm 0.02$ | 1.9 | $> 4$ | $-28.6 \pm 0.5$ | $6.6 \pm 0.6$ |
| $\chi$ Cyg | $4.7 \pm 0.1$ | 63 | $> 4$ | $7.8 \pm 0.2$ | $9.9 \pm 0.2$ |
| (4–3) | $4.0 \pm 0.07$ | 52.7 | 1.4 | $9.9 \pm 0.1$ | $9.0 \pm 0.1$ |
| Z Cyg | $0.27 \pm 0.02$ | 1.5 | $> 4$ | $-147.7 \pm 0.2$ | $4.0 \pm 0.3$ |
| T Cep | $1.5 \pm 0.07$ | 9.4 | $> 4$ | $-3.0 \pm 0.2$ | $4.8 \pm 0.3$ |
| R Peg | $0.23 \pm 0.02$ | 2.5 | $> 4$ | $22.7 \pm 0.5$ | $8.3 \pm 0.8$ |
| W Peg | $0.31 \pm 0.04$ | 3.6 | 1.2 | $-15.5 \pm 0.5$ | $7.7 \pm 0.9$ |
| R Cas | $4.7 \pm 0.04$ | 87 | 0.9 | $23.7 \pm 0.0$ | $12.1 \pm 0.2$ |
| (4–3) | $6.0 \pm 0.29$ | 102.4 | $> 4$ | $25.6 \pm 0.2$ | $13.3 \pm 0.3$ |

[A] The CO(4–3) profile of o Ceti not well fit by equation 1 for any combination of parameters (see figure 2).

parameters. These confidence limits were derived from the covariance matrix produced by the nonlinear least–squares program which was used to derive the parameters; they are merely formal errors, and do not include such effects as telescope pointing errors etc. The values of $V_o$ have been corrected for the broadening caused by the 0.9 km s$^{-1}$ resolution of the 500 MHz AOS, by assuming $V_o$ and the velocity resolution add in quadrature. No such correction was applied to the CO(4–3) line fits (except for that of R Cas), because the more narrow 50 MHz AOS used for those observations did not broaden the lines measurably. Figures 1a–d and 2a–b show the CO(3–2) and CO(4–3) spectra of the detected stars, along with the fitted profiles.

Three of the stars; T Ari, R Cnc and RR Sco, have outflow velocities of less than 4 km s$^{-1}$, lower than any Mira variable previously detected in CO (to the author's knowledge). The average outflow velocity is only 7.0 km s$^{-1}$. This is substantially below the value found in previous surveys in which candidates were selected by infrared criteria. For example the survey by Nyman *et al.* (1992) used *IRAS* color criteria to select candidates. Their survey detected 20 oxygen–rich Miras, with an average outflow velocity of 13.3 km s$^{-1}$. Zuckerman and Dyck (1986a, 1986b, 1989, Zuckerman, Dyck & Claussen 1986) surveyed the evolved stars with the largest fluxes in the AFGL and *IRAS* catalogs; among their detections were 14 oxygen–rich Miras with an average outflow velocity of 11.3 km s$^{-1}$.

### 4.1 Detections and Nondetections

*IRAS* PSC positions were used for 24 of the 66 survey stars. Even for stars which are bright in the infrared the PSC positions have a mean error of $\sim 9$" (Beichman *et al.* 1988), a sizable fraction of the CSO's 20" CO(3–2) beam. As can be seen in table 1, the detection rate for objects for which *IRAS* coordinates were used was not high. However the estimated distances for these objects is 50% greater on average than those of the stars with SAO coordinates. It is likely that the low detection rate for the *IRAS* coordinate stars is due at least in part to their greater average distance. If we consider only those stars more distant than 300 pc, there was not a dramatic difference between the detection rate for the SAO coordinate stars (46% detected) and the stars for which *IRAS* positions were used (44% detected). Nonetheless any individual non-detection must be viewed with suspicion if its position was taken from the PSC.

Figure 1a

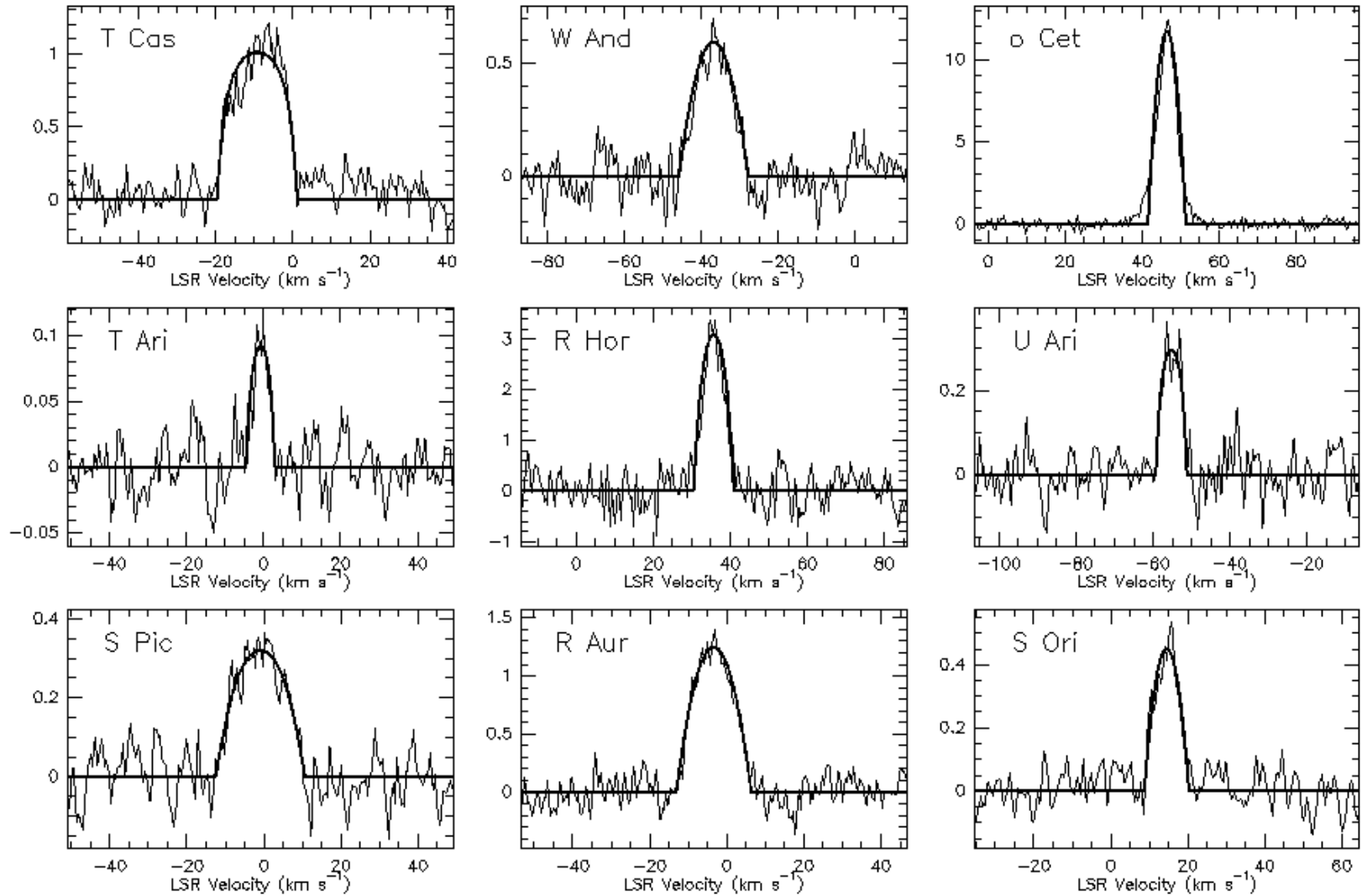

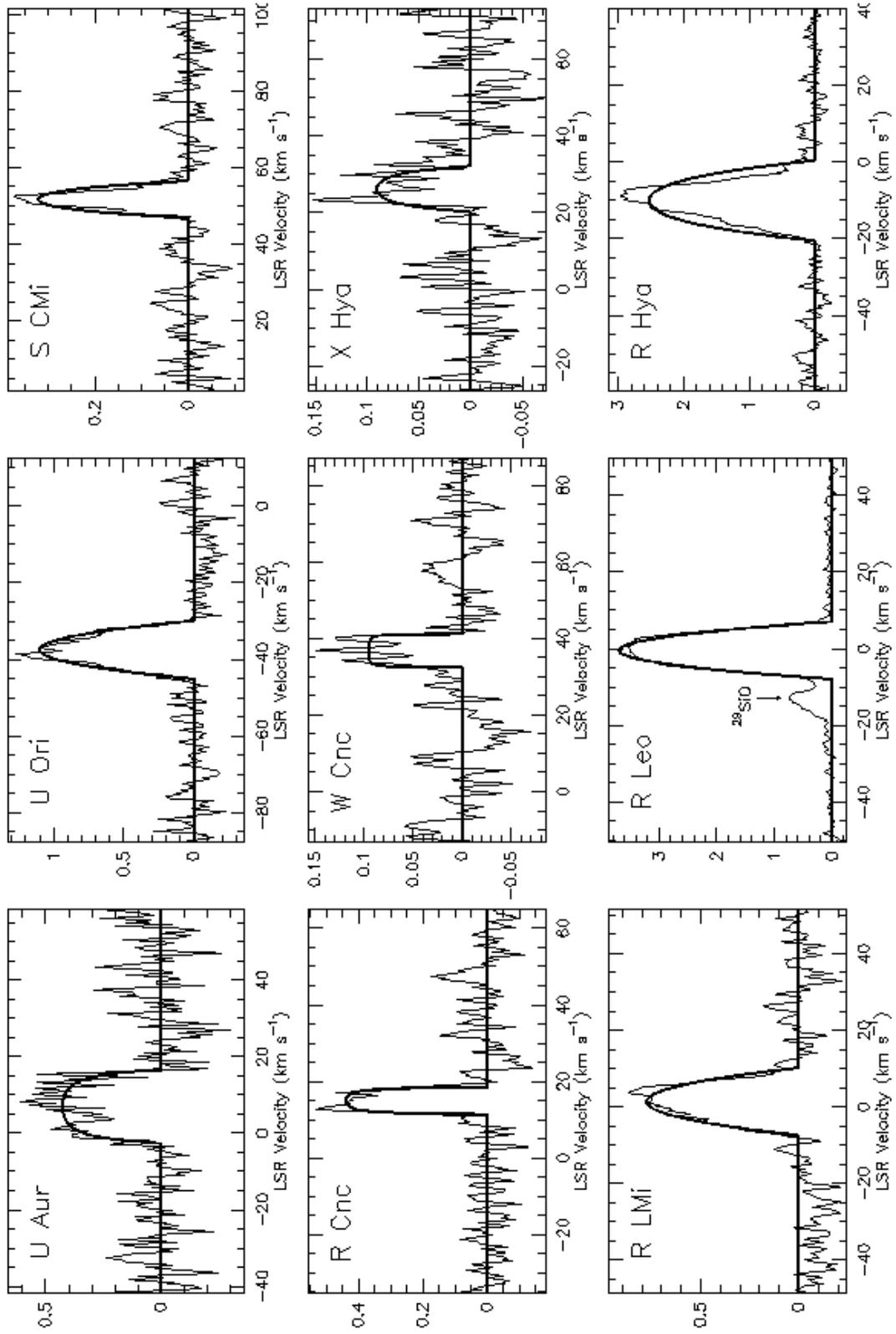

Figure 1b

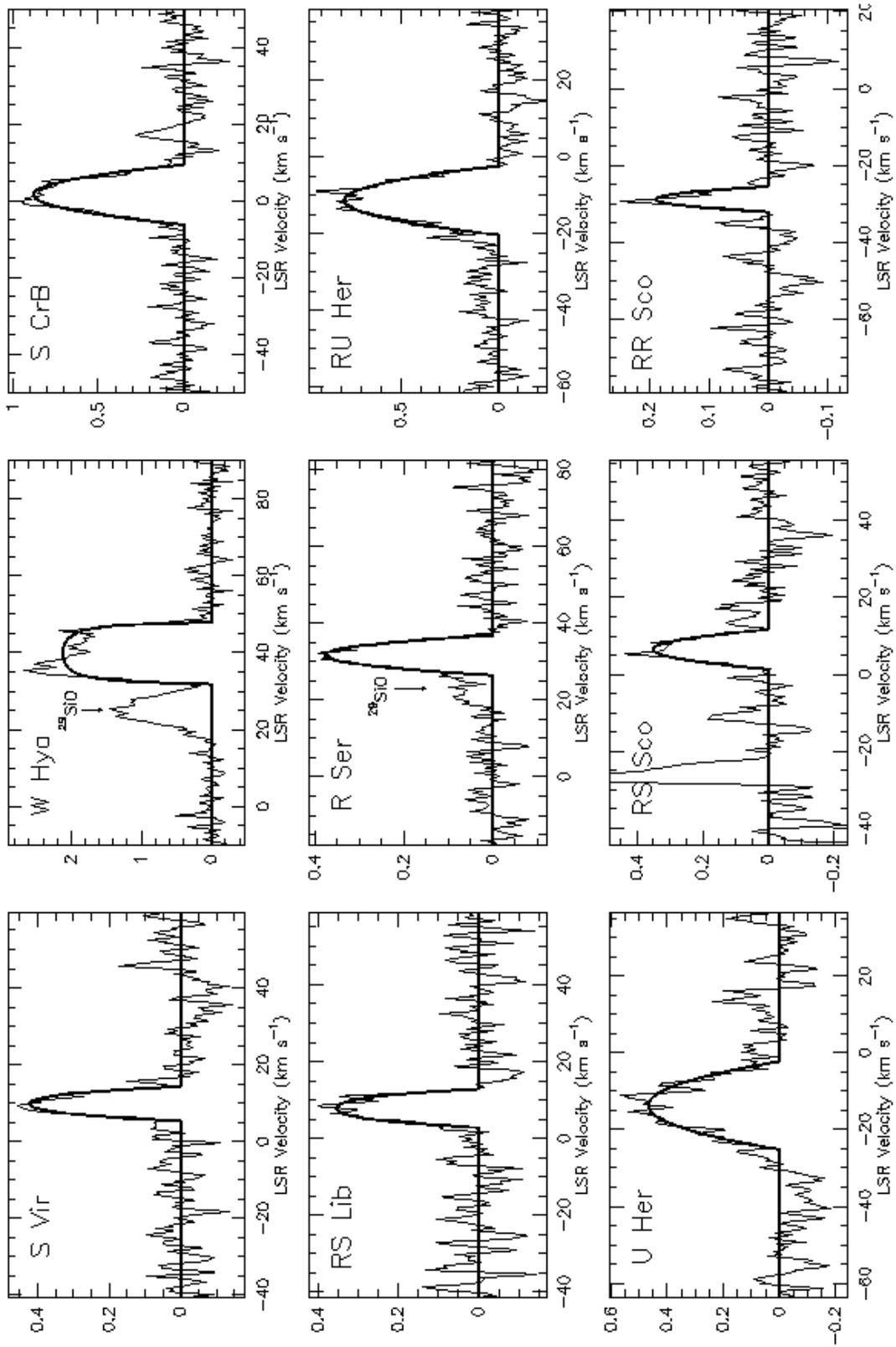

Figure 1c

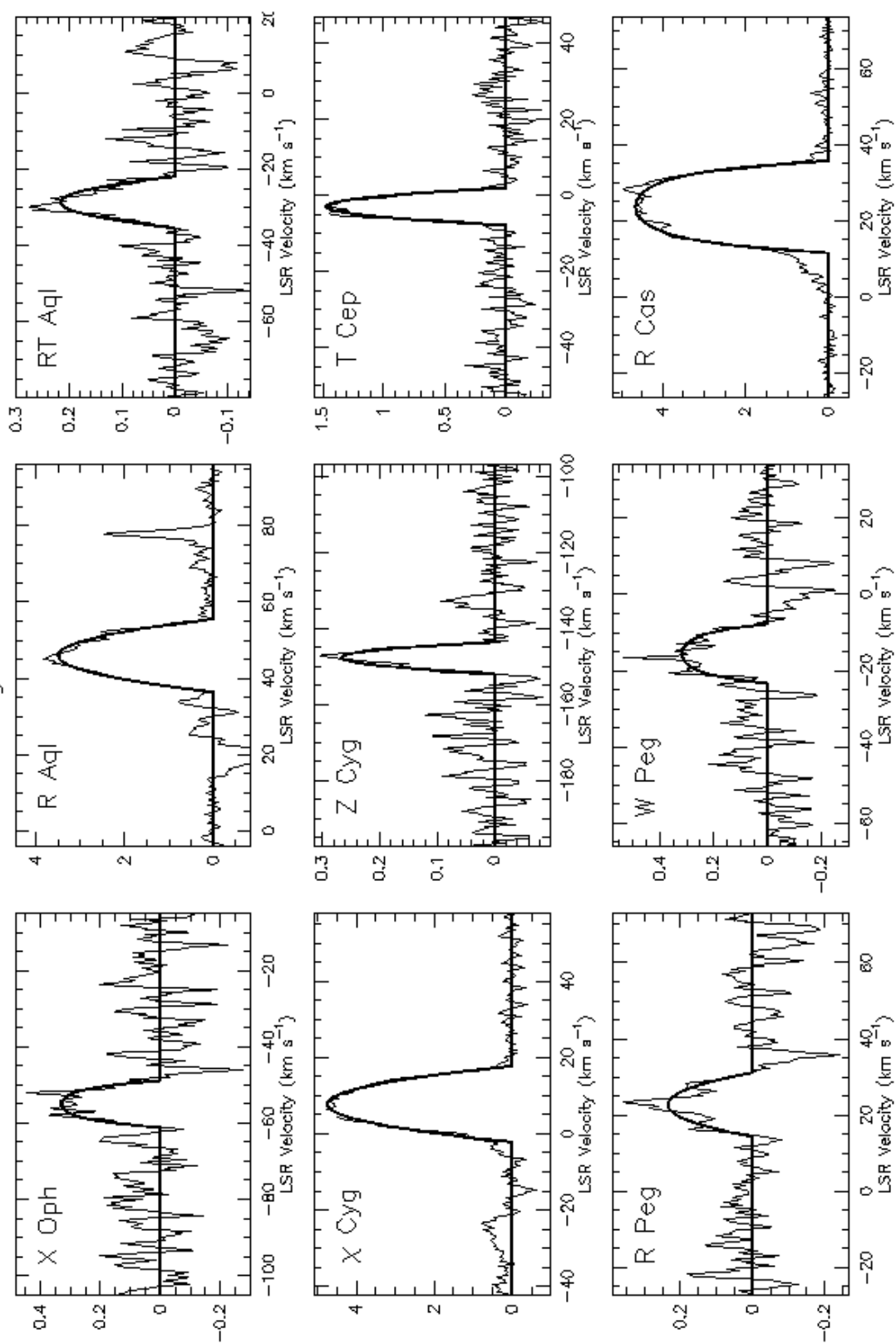

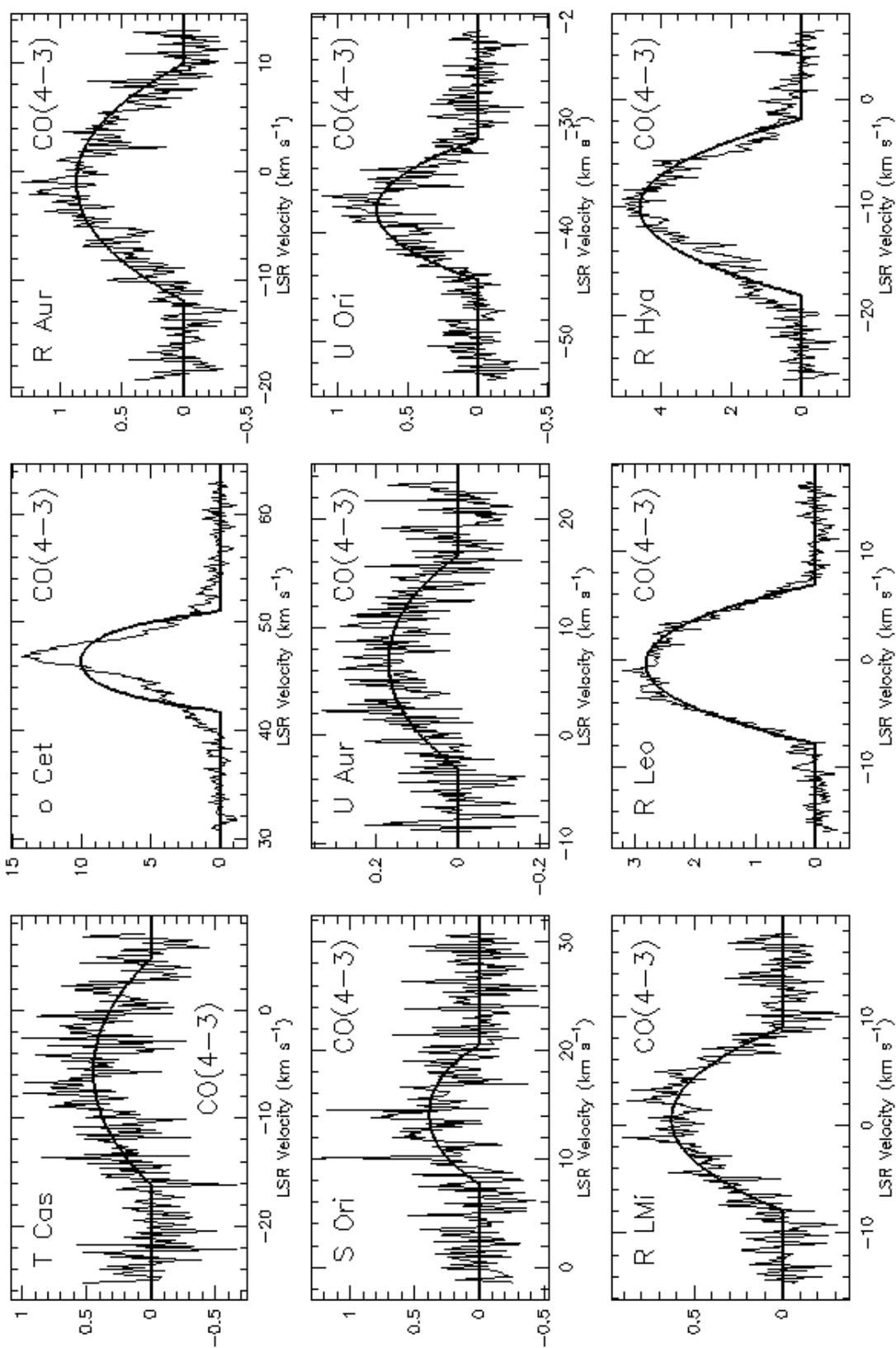

Figure 2a

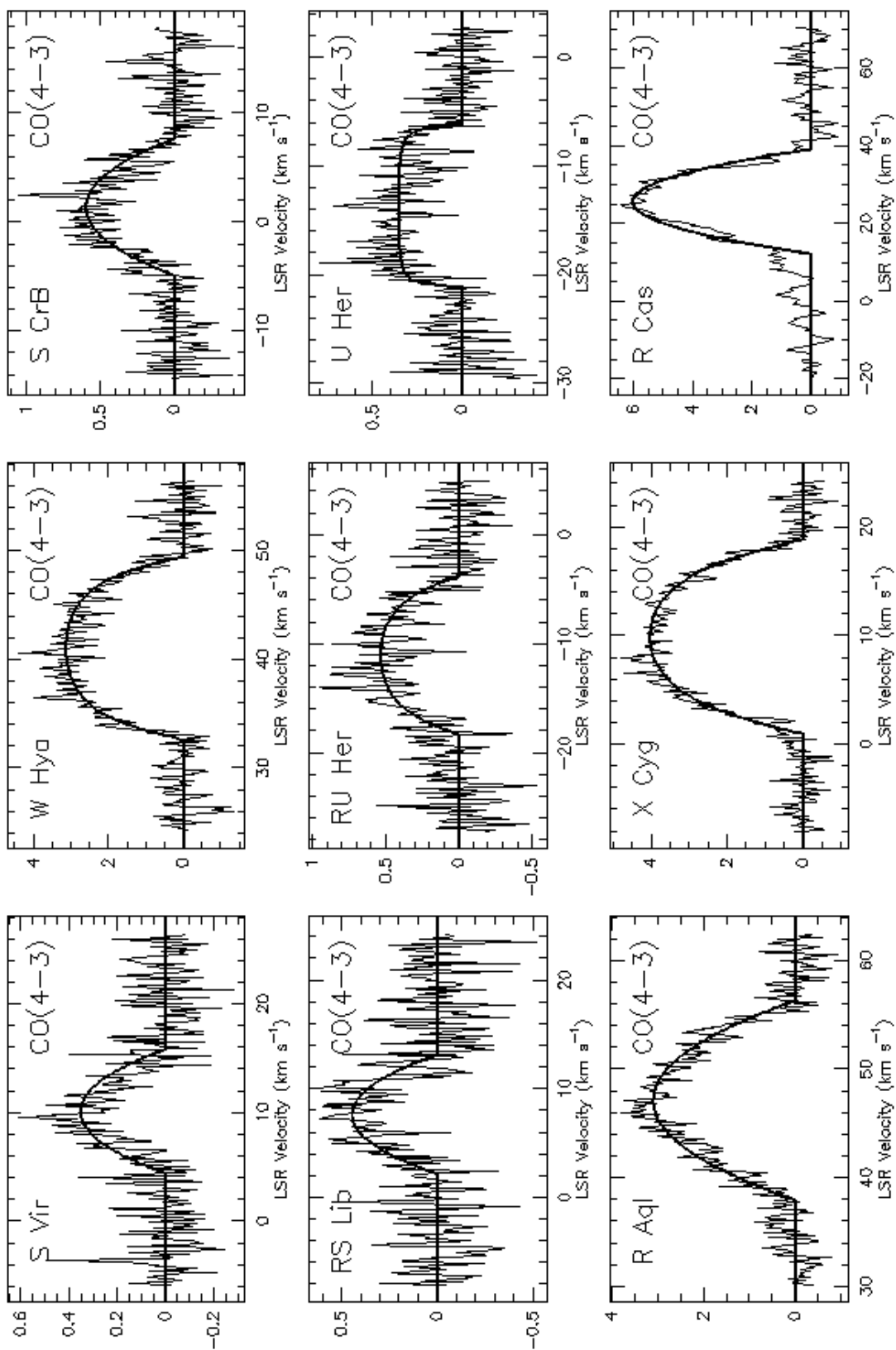

Figure 2b

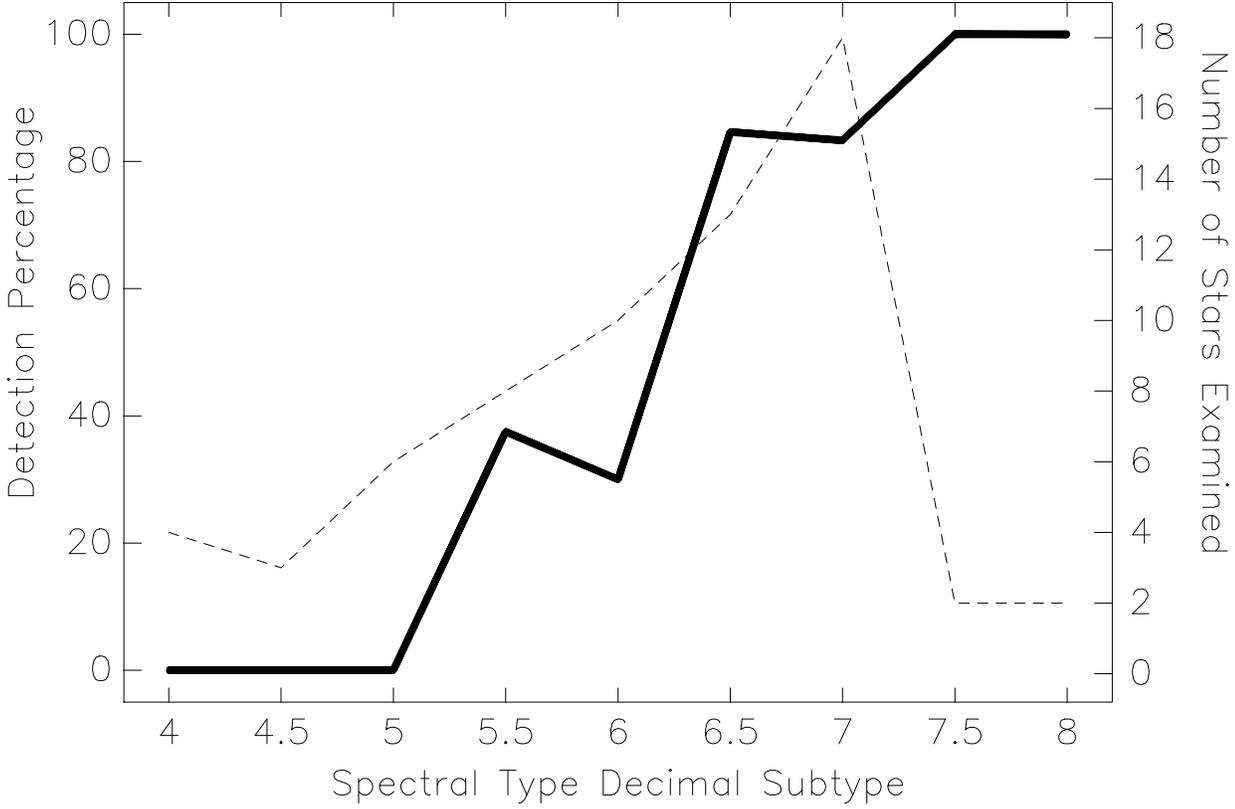

**Fig. 3.** The horizontal axis denotes the decimal subtype of the star's spectral type at visual maximum. A value of 5.5 could denote a spectral type of either M5.5 or S5.5. The bold line shows the percentage of stars detected. The dashed line shows the number of stars examined for each spectral type.

The characteristic that best discriminates between the stars which were detected and those which weren't is the spectral type's decimal subtype (Yerkes system), taken from the RCMV. As figure 3 shows, the detection percentage is a steep and nearly monotonic function of the subtype. None of the 13 stars examined with subtypes below 5.5 were detected, while all 4 with subtypes greater than 7 were. An attempt was made to find an infrared color index which divided the stars cleanly into detected and nondetected groups. All colors of the form

$$C = \frac{f_I^{\lambda(1)} f_K^{\lambda(2)} f_{4\mu}^{\lambda(3)} f_{11\mu}^{\lambda(4)} f_{12\mu}^{\lambda(5)} f_{25\mu}^{\lambda(6)} f_{60\mu}^{\lambda(7)} f_{100\mu}^{\lambda(8)}}{f_I^{\lambda(9)} f_K^{\lambda(10)} f_{4\mu}^{\lambda(11)} f_{11\mu}^{\lambda(12)} f_{12\mu}^{\lambda(13)} f_{25\mu}^{\lambda(14)} f_{60\mu}^{\lambda(15)} f_{100\mu}^{\lambda(16)}}$$

were examined, where $f_I$ and $f_K$ are the I and K band fluxes from the TMSS catalog, $f_{4\mu}$ and $f_{11\mu}$ are from the AFGL catalog, the four long wavelength fluxes are from the *IRAS* PSC and $0 \leq \lambda(n) \leq 2$. No color of this form was found to be able to predict which stars would be detectable in all cases, although several of these colors were nearly able to do so. Also, no color was a more reliable indicator of detectability than the *IRAS* 25$\mu$ flux alone. Three of the undetected stars; R Aqr, R Car and R Cen, were consistently included among the detected stars by all the colors and fluxes which most reliably discriminated between detectable and not detectable sources. R Aqr is a symbiotic variable, and this might explain its lack of detectable CO emission. R Cen and R Car are normal Miras however, and the lack of CO surrounding these stars is intriguing. Both have larger fluxes in all four *IRAS* bands than many of the stars which were detected, indicating that they do have substantial dust shells. As neither of these stars rise above 10 degrees from the southern horizon, it's tempting to dismiss their nondetection as a result of poor instrument performance at low elevations. However Nyman *et al.* (1992) also observed R Car from the southern hemisphere (at SEST), and they too did not detect it. The *V* band amplitudes of R Car and R Cen, 6.0 and 6.5 mag. respectively (Eggen 1992), are fairly low for Mira variables, but they are as large or larger than several of the detected stars. However at visual band maximum the spectral types of R Car and R Cen, M5e and M4.5e respectively (RCMV), are earlier than any of the detected objects. Perhaps the hotter photospheres of these two stars prevents the dust forming near enough to the star to allow efficient dust–gas momentum coupling and mass loss. Alternatively, these stars may be losing mass at rates comparable to cooler Miras, but carbon monoxide might not be fully associated in their photospheres. It would be surprising if R Car and R Cen had strong winds with no CO, because a CO envelope has been detected surrounding $\alpha$ Ori (Knapp *et al.* 1982) which has an even warmer photosphere (M2Iab).

**Table 3. – Calculated Luminosities, Distances and Mass Loss Rates**

| Star | $M_v$ | Distance (pc) | $\dot{M}$ ($10^{-8}$ M$_\odot$ yr$^{-1}$) | $R_{Max}$ ($10^{16}$ cm) | Star | $M_v$ | Distance (pc) | $\dot{M}$ ($10^{-8}$ M$_\odot$ yr$^{-1}$) | $R_{Max}$ ($10^{16}$ cm) |
|---|---|---|---|---|---|---|---|---|---|
| T Cas | -0.54 | 350 | 110 ± 36 | 16. | S Vir | -1.87 | 450 | 5.6 ± 2 | 5.6 |
| (4-3) | | | 45 ± 17 | | (4-3) | | | 5.7 ± 2 | |
| W And | -0.16 | 280 | 34 ± 12 | 9.7 | W Hya | -0.12 | 80 $^A$ | 18 ± 7 | 7.5 |
| o Cet | -1.84 | 100 $^A$ | 36 ± 12 | 14. | (4-3) | | | 21 ± 8 | |
| (4-3) | | | 21 ± 8 | | S CrB | -0.83 | 310 $^A$ | 35 ± 12 | 11. |
| T Ari | -1.63 | 350 | 0.6 ± 0.2 | 2.4 | (4-3) | | | 11 ± 4 | |
| R Hor | -0.98 | 190 | 28. ± 11 | 12. | RS Lib | -0.11 | 210 | 3.7 ± 1 | 3.5 |
| U Ari | -1.44 | 710 $^A$ | 6.5 ± 2 | 6.3 | (4-3) | | | 3.5 ± 1 | |
| S Pic | -0.77 | 440 | 59 ± 24 | 11. | R Ser | -1.35 | 330 $^A$ | 4.5 ± 1 | 4.7 |
| R Aur | -1.11 | 450 | 160 ± 60 | 20. | RU Her | -0.89 | 410 | 75 ± 24 | 14. |
| (4-3) | | | 110 ± 40 | | (4-3) | | | 28 ± 10 | |
| S Ori | -0.75 | 480 | 21 ± 7 | 9.4 | U Her | -0.72 | 360 | 59 ± 24 | 11. |
| | | | 22 ± 8 | | (4-3) | | | 14 ± 5 | |
| U Aur | -0.71 | 500 | 63 ± 24 | 13. | RS Sco | -1.03 | 300 | 2.2 ± 0.8 | 4.3 |
| (4-3) | | | 24 ± 9 | | RR Sco | -0.96 | 200 | 0.6 ± 0.2 | 2.2 |
| U Ori | -0.90 | 240 | 28 ± 10 | 9.5 | X Oph | -0.80 | 270 | 4.5 ± 1 | 4.2 |
| | | | 9.1 ± 3 | | R Aql | -0.67 | 190 | 110 ± 36 | 17. |
| S CMi | -0.52 | 340 | 4.1 ± 1 | 4.6 | (4-3) | | | 76 ± 30 | |
| R Cnc | -0.88 | 300 | 2.3 ± 0.8 | 4.0 | RT Aql | -0.52 | 450 | 10 ± 4 | 6.2 |
| W Cnc | -1.61 | 720 $^A$ | 3.1 ± 1 | 4.3 | χ Cyg | -0.29 | 115 $^C$ | 82 ± 24 | 14. |
| X Hya | -0.16 | 400 | 3.0 ± 1 | 3.6 | (4-3) | | | 46 ± 20 | |
| R LMi | -0.47 | 260 $^A$ | 36 ± 12 | 10. | Z Cyg | -1.11 | 490 | 4.0 ± 1 | 5.0 |
| (4-3) | | | 23 ± 9 | | T Cep | -0.94 | 220 | 9.9 ± 4 | 7.2 |
| R Leo | -0.46 | 120 $^B$ | 41 ± 12 | 12. | R Peg | -0.74 | 380 $^A$ | 12 ± 5 | 6.1 |
| (4-3) | | | 25 ± 9 | | W Peg | 0.03 | 270 $^A$ | 9.8 ± 4 | 5.6 |
| R Hya | -0.76 | 110 $^A$ | 42 ± 12 | 10. | R Cas | -0.78 | 230 | 290 ± 110 | 24. |
| (4-3) | | | 39 ± 20 | | (4-3) | | | 360 ± 100 | |

$^A$Distance from Jura and Kleinmann 1992, $^B$Distance from Gatewood 1992

$^C$Distance from Stein 1991

## 4.2 Mass Loss Rates

After the line profile parameters had been calculated, an artificial CO(3–2) profile was calculated using a Large Velocity Gradient (LVG) radiative transfer program (Morris 1980). The mass loss rate was adjusted until the peak $T_{mb}$ of the artificial profile matched that of the parameterized profile.

The CO envelope was assumed to be truncated by photodissociation, and this radius was calculated using the photodissociation rate of Mamon *et al.* (1988). Table 3 lists the calculated absolute magnitudes, distances, mass loss rates ($\dot{M}$) and the outer radii of the CO envelopes for the stars which were detected. The highest (R Cas: $2.9 \times 10^{-6} M_\odot \mathrm{yr}^{-1}$) and lowest (T Ari and RR Sco: $6 \times 10^{-9} M_\odot \mathrm{yr}^{-1}$) mass loss rates differ by nearly 3 orders of magnitude. The star's distance, luminosity and fractional abundance of CO were input parameters to the LVG model, and the sensitivity of the calculated value of $\dot{M}$ to errors in these parameters was investigated by varying the parameters by their estimated $1\sigma$ uncertainties, and recalculating $\dot{M}$.

Errors in the star's estimated luminosity and distance are not independent, because the distance was calculated using the luminosity provided by the period – spectral type – luminosity relationship (PSLR). If a particular star is actually more luminous than indicated by the PSLR, it provides more infrared radiation to excite CO, which leads to increased CO(3–2) radiation from a given amount of CO. Increasing the estimated luminosity of the star decreases the value of $\dot{M}$ calculated by the LVG model, because less molecular material is needed to produce the same line strength. However if the star's luminosity has been underestimated, then its derived distance will also be too low. Increasing the distance estimate given to the LVG increases the calculated value of $\dot{M}$, because less of the CO(3–2) radiation from a given amount of CO will arrive at the telescope. Therefore the errors in $\dot{M}$ introduced by errors in the luminosity and distance will have opposite signs. The net effect is that varying the luminosity by $\pm 0.4$ magnitudes, the mean spread in the PSLR (Celis 1981), changes the derived distances by (+20%,–17%) and $\dot{M}$ by $\pm 23\%$, on average. The $\dot{M}$ values shown in Table 3 were calculated with the assumption $f = [\mathrm{CO}]/[\mathrm{H}_2] = 3 \times 10^{-4}$, as would be the case if hydrogen and carbon were present in the solar abundance ratio, and half of the carbon was incorporated into CO molecules. Decreasing $f$ to $2 \times 10^{-4}$ increases $\dot{M}$ by 39% on average, while increasing $f$ to $4 \times 10^{-4}$ on average decreases $\dot{M}$ by 21%. If errors in the estimated luminosity and $f$ are independent, then they should add in quadrature, and the uncertainty in the calculated value of $\dot{M}$ arising from these two sources is $\pm 35\%$. Error estimates for the value of $\dot{M}$ for individuals stars, using the range of parameters discussed above, are given in Table 3.

Figure 4 shows a log–log plot of the calculated mass loss rates versus the outflow velocity. There's an obvious strong correlation, indicating a power–law relationship holds between these quantities. The best–fit power–law is

$$\dot{M} \approx 2.94 \times 10^{-10} \; V_{o \; (\mathrm{km\,s^{-1}})}^{3.35} \; M_\odot \mathrm{yr}^{-1}. \qquad (2)$$

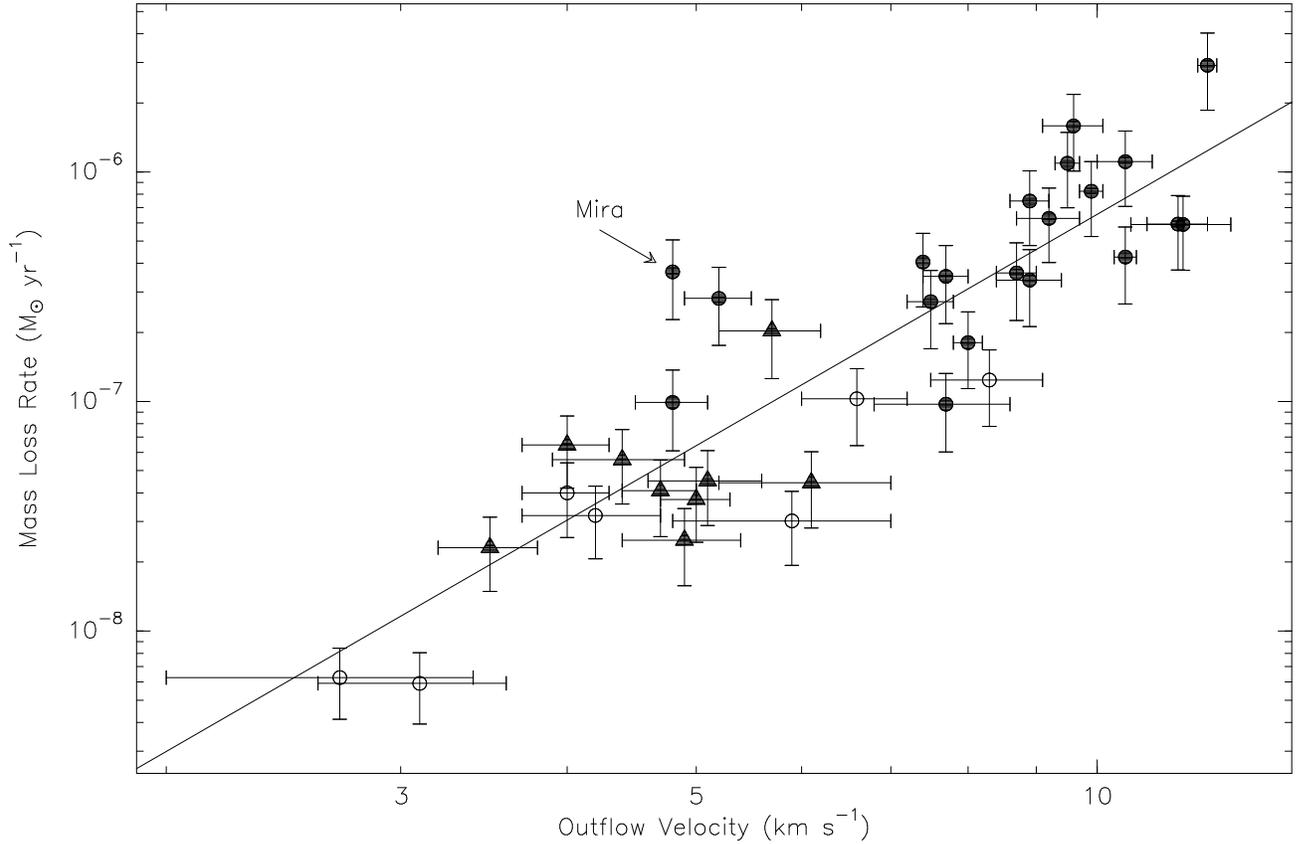

**Fig. 4.** A Log-Log plot of mass loss rate as a function of outflow velocity is shown for the 36 Mira stars which were detected. The line shown is the linear least-squares best fit to the data (equation 2). Filled circles indicate stars with integrated CO(3–2) line intensities of 3 K km s$^{-1}$ or greater, which lie above line D in figure 5. Open circles represent stars which fall below line C in figure 5.

Ironically, the star which deviates most strongly from this relationship is Mira itself! The most likely reason for this is that Mira's true outflow velocity is substantially larger than the 4.8 km s$^{-1}$ found by fitting equation 1 to the data. The wings of Mira's CO(3–2) profile extend to about 10 km s$^{-1}$ from the line center (Knapp 1995). Plotting Mira in figure 4 with a higher expansion velocity would move it closer to the locus of points, but the peculiar high velocity wings on both the 3–2 and 4–3 profiles indicate that the simple spherical, constant expansion velocity LVG model used to calculate Ṁ is too simplistic for this star.

It is important to recognize that selection effects could influence the appearance of figure 4 and hence equation 2. Figure 5 shows the integrated intensity as a function of expansion velocity, on a log–log plot. If we assume that the weakest lines are unresolved and optically thin, then the line area is approximately given by twice the product of the peak antenna temperature (T), and $V_o$. If the spectrometer is well behaved, the noise in the measurement of the line area will be proportional to $V_o^{1/2}$, so the signal/noise for a weak line will be proportional to $TV_o^{1/2}$. For a fixed signal/noise detection threshold, the minimum detectable antenna temperature, $T_{min}$, should therefore be proportional to $V_o^{-1/2}$, and the minimum detectable line area should be $T_{min}V_o^{1/2}$. Thus one would expect the points of figure 5 to be bounded by a line with a slope of 1/2. However, on a more subjective level, very wide, weak lines are apt to be overlooked or incorrectly attributed to instrument baseline irregularities, leading to false *non*detections. In this way one could end up with a list of detections in which all the least intense lines, which for a given distance indicate a low mass loss rate, have small outflow velocities. This is worrisome, because such a bias might lead to a relationship resembling equation 2. If all lines above some threshold temperature were detected regardless $V_o$, then one would expect the points to be bounded by a line with a slope of 1. If broad, weak lines were overlooked, then one would expect the bounding line to have a slope greater than 1. In fact, the points all lie on or above line A in figure 5, which has a slope of 0.7, suggesting that weak, broad lines were not systematically rejected. However, there are no points near line A with large expansion velocities, and if one star (X Hya) is rejected, the points seem to be bounded more closely by line B, which has a slope of 1.8. So the sample is probably too small to allow the importance of selection effects to be quantitatively assessed. A reasonable guess would be that the 7 points falling below line C on figure 5, which has a slope of 1, represent stars with antenna temperatures which might not have been recognized as detections, had their outflow velocities been higher. In figure 4, these 7 stars are represented by open circles, and it is clear that the line produced by equation 2 is still a good fit to the data even if these points are rejected. Above 3 K km s$^{-1}$ (line D in figure 5), there is no correlation at all between line area and outflow velocity. Figure 4 shows that even if only those points lying within this uncorrelated region of figure 5 are used, there is still a correlation between M and V$_o$ .

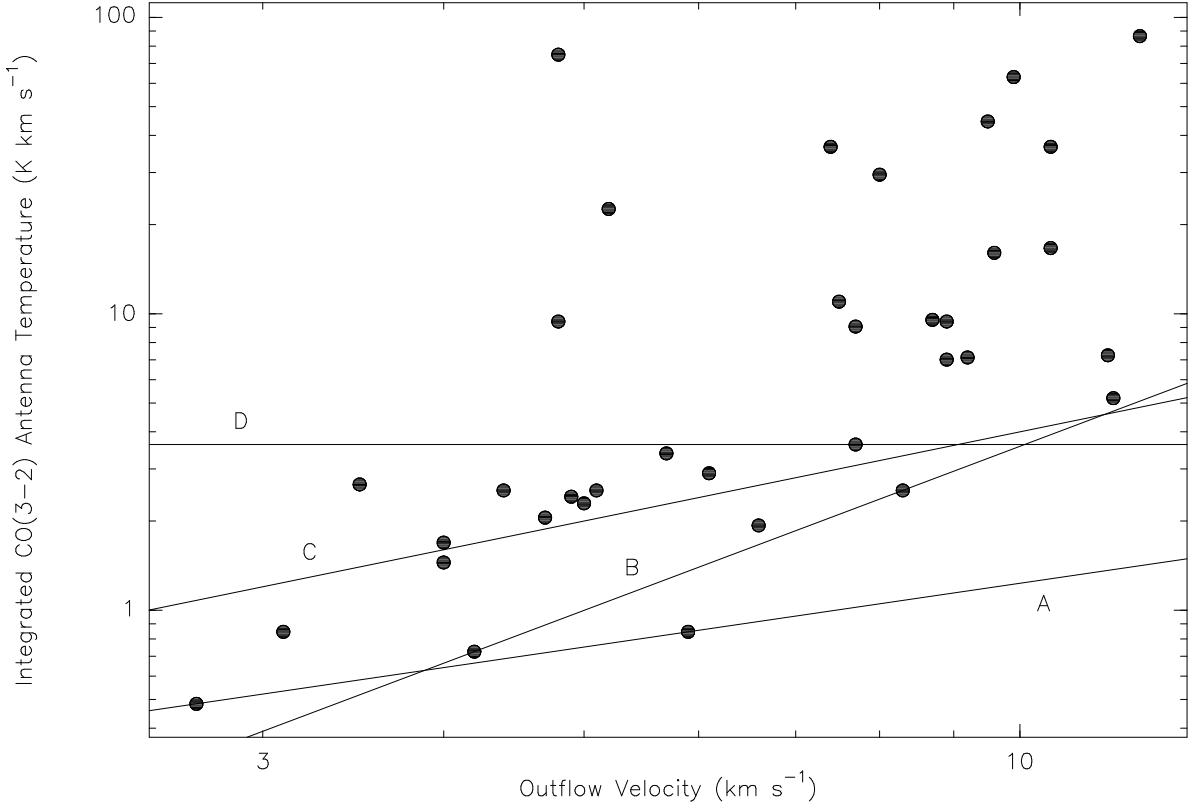

**Fig. 5.** The CO(3-2) line area, as a function of outflow velocity, is shown. Above 3 K km s$^{-1}$ there is no correlation.

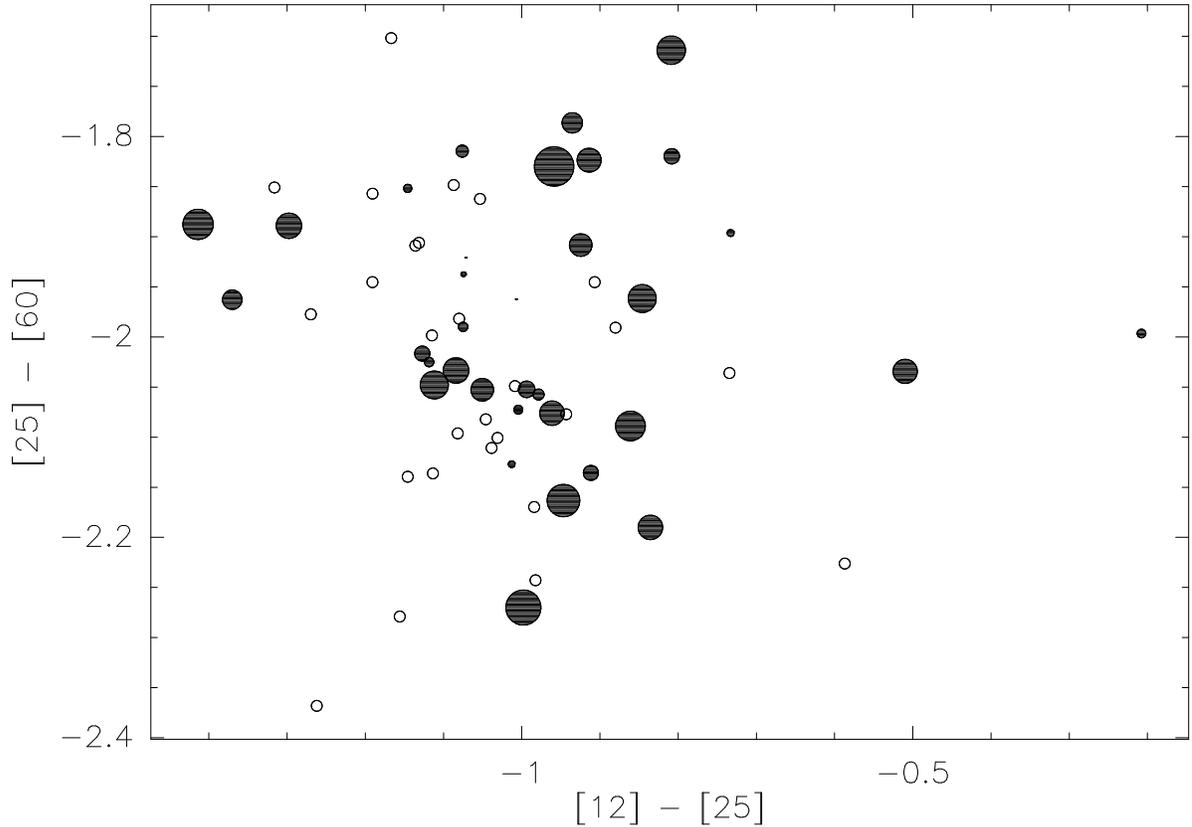

**Fig. 6.** An IRAS 2-color diagram, for the detected and nondetected stars is shown. The abscissa is $2.5\log_{10}(F_{25\mu}/F_{12\mu})$ and the ordinate is $2.5\log_{10}(F_{60\mu}/F_{25\mu})$. Open circles represent nondetections, filled circles represent detections. The size of the filled circles is proportional to the log of the mass loss rate listed in table 3. Five survey stars with poor quality 12, 25, or $60\mu$ fluxes in the PSC (RS Mon, R Cen, T Nor, RS Sco and R Aqr) have not been included in this plot.

Van der Veen and Habing (1988) showed that oxygen rich Miras and OH/IR stars fall along a broad track on an *IRAS* 2-color diagram constructed from the PSC 12, 25 and $60\mu$ fluxes. They further showed that as one moves along this path from the warmer to the cooler end, stars with increasingly large mass loss rates are found. These authors interpret this as an evolutionary progression. Figure 6 shows such a 2-color diagram for the stars in this survey. Surprisingly, the detected and nondetected stars are not well segregated and the mass loss rate seems unrelated to either of the colors. Though not shown in figure 6, colors involving the *IRAS* 100$\mu$ flux were also examined and they too appeared uncorrelated with the mass loss rate. It has been shown that for highly obscured stars the mass loss rate is strongly dependent on IRAS color, and empirical formulae have been constructed to give mass loss estimates based on these colors (van der Veen 1989). The results presented here show that such formulae are not applicable to stars with visible photospheres and low values of Ṁ. On the other hand, figure 7 shows that the mass loss rates in table 3 do correlate with all four *IRAS* fluxes if they are normalized to a common distance, although the correlation with 100$\mu$m flux is very weak. The correlation is strongest with the 60$\mu$ flux. Such a correlation should be viewed with caution, because errors in the stellar distance estimate influence both the calculated mass loss rate and the distance normalized flux in the same manor, and can produce a spurious correlation. However neither Ṁ nor the 60$\mu$m luminosity shows any correlation with distance (see figures 7E and 7F), therefore it seems likely that the correlation of Ṁ with infrared luminosity is real.

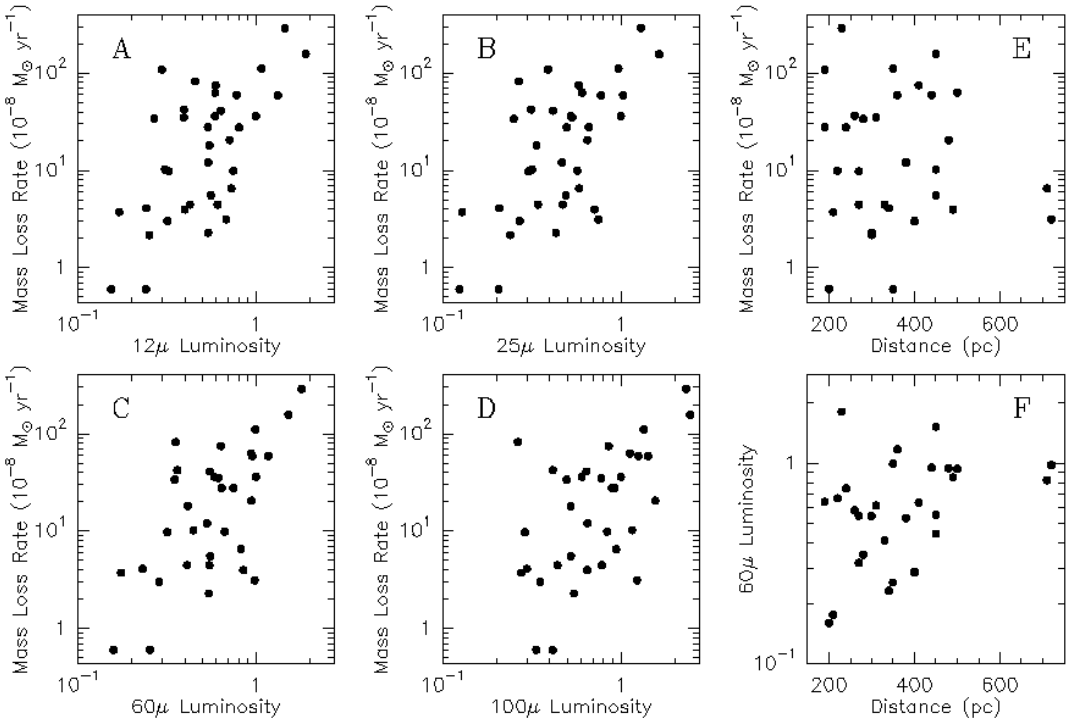

**Fig. 7.** Figures **A** through **D** show the mass loss rate as a function of far infrared luminosity (calculated as flux × distance$^2$) plotted for each of the *IRAS* bands. Stars with poorly determined *IRAS* fluxes are not included. In each case, the luminosity has been normalized to that of o Ceti. Figures **E** and **F** show there is no correlation between distance and the calculated mass loss rates and 60$\mu$m luminosity, implying that these quantities aren't dominated by the effects of errors in the distance estimates.

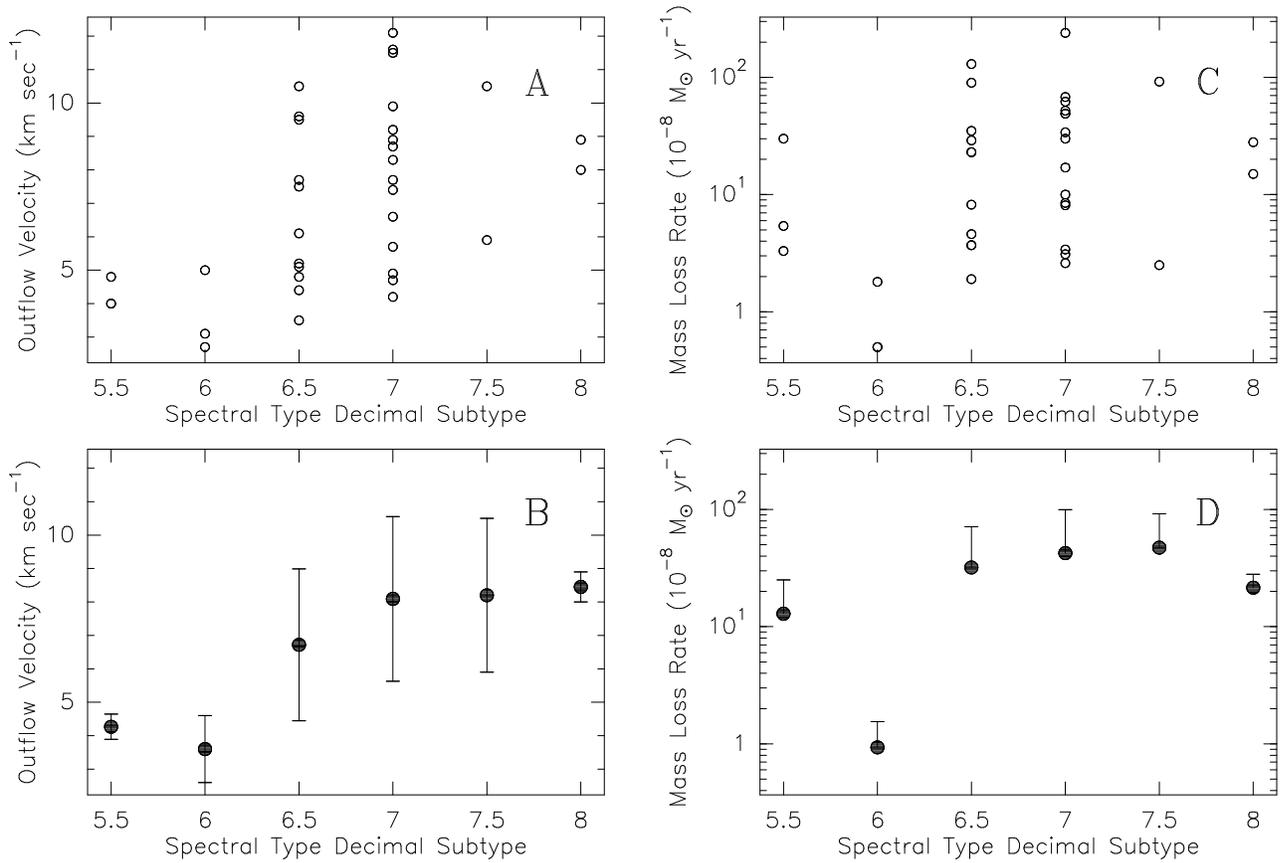

**Fig. 8.** (**A**) shows the distribution of CO(3–2) outflow velocities as a function of decimal spectral subtype at $V$ band maximum. (**B**) shows the averages of the points in plot A. The error bars indicate the standard deviation. (**C**) shows the mass loss rates from table 3 as a function of decimal subtype, and (**D**) once again shows the averages.

## 4.3 The Wind Velocity

There is general agreement that the copious mass loss from heavily dust enshrouded stars, such as the OH/IR stars, is driven by radiation pressure on dust grains. However there is no such consensus on the cause of the much more modest mass loss seen in visible Miras. Netzer and Elitzur (1993) solved the equations of motion for dust and gas in a radiation driven wind, and concluded that there is a lower limit of $\sim 10^{-7}$ $M_\odot \mathrm{yr}^{-1}$ for radiation driven mass loss, well above many of the values of $\dot{M}$ in table 3. Fusi–Pecci and Renzini (1975) showed that acoustic energy could drive mass loss in red giant stars. Yet the increase in outflow velocity at later spectral types shown in figures 8A&B is what would be expected in a radiation driven wind. Netzer and Elitzur (1993) show that for a such a wind, if the terminal velocity of the gas greatly exceeds the initial velocity that the gas has at the base of the acceleration region, then $V_o \propto R_o^{-1/2}$ where $R_o$ is the distance at which dust forms. Thus cooler stars, which should allow dust to form at smaller radii, will have higher values of $V_o$. The absence of detectable CO emission from the stars with spectral types earlier than M5.5 indicates that either such stars have extremely low mass loss rates or that they have mass loss rates similar to the cooler stars, but the gas is atomic. Knapp *et al.* (1995) looked for the 609$\mu$m line of C I in the envelope of Mira (spectral type M5.5) but did not detect it. Thus it would be surprising if large amounts of atomic carbon would be seen surrounding Miras with only slightly warmer photospheres. It seems more likely that these warmer Miras have very low mass loss rates. These warm Miras may inhibit the formation of dust throughout the region where the gas density is high enough to support a significant mass loss rate. It therefore appears that the results of this survey support radiation pressure on grains as the driving mechanism for even low mass loss rates. It is surprising however, that $\dot{M}$ is not correlated with spectral type (figures 8C&D).

One of the strongest pieces of evidence that the winds from AGB stars are driven by radiation pressure is the fact that the momentum present in the wind is almost always less than or comparable to the momentum available in the radiation field (Knapp *et al.* 1982). Assuming radiation pressure drives the wind, the rate of increase of the wind's mechanical momentum ($\dot{M}V_o$) can exceed the momentum available from the star's radiation ($L_*/c$) only if the light is scattered in an optically thick dust envelope. All of the stars in this survey must have optically thin dust envelopes, because the photospheres are still visible. If we assume that all Miras have a luminosity of about 3400 $L_\odot$ (the average of the oxygen Miras in the LMC survey of Feast *et al.* 1989), and if we use equation 2 to estimate the mass loss rate, then there is a maximum value of $V_o$, $V_{max}$, for which $\dot{M}V_o < L_*/c$. This value is 17 km s$^{-1}$. Because equation 2 is such a strong function of $V_o$, $V_{max}$ is rather insensitive to the assumed Mira luminosity; if the true luminosities of Mira variables ranged from $10^3$ to $10^4$ $L_\odot$, this would only introduce a range of $\pm40\%$ in $V_{max}$. It seems therefore that any AGB star losing mass by the same mechanism as these Miras will be obscured by an optically thick envelope if $V_o$ is $\gtrsim$ 17 km s$^{-1}$.

In the standard radiation driven model of the wind, the dust feels the radiation pressure, and propels the gas outward via collisions. Thus the dust must be moving though the gas, at a velocity called the "drift velocity," for which Goldreich and Scoville (1976) derived the expression

$$v_d \approx \left( \frac{Q L_* V_o}{\dot{M} c} \right)^{1/2} \qquad (3)$$

where $Q$ is the momentum coupling constant between the radiation field and the dust. If we assume that the dust consists of silicate spheres with radii of $0.1\mu$, and the radiation field is that of a 2000 K black body, equation 3 becomes

$$v_d \approx \left( \frac{3 Q_{ext} L_* V_o}{2 \dot{M} c} \right)^{1/2} = \left( \frac{3 \times 0.023 \times 3400 L_\odot \times V_o}{2 \dot{M} c} \right)^{1/2}$$

where $Q_{ext}$, the dust extinction efficiency, has been taken from a tabulation by Draine (1987). Substituting equation 2 for $\dot{M}$ gives

$$v_d \approx 61 \times V_o^{-1.18} \approx \frac{61}{V_o} \quad \text{km s}^{-1} \; . \qquad (4)$$

At $V_o = 3$ km s$^{-1}$, this gives a drift velocity of 20 km s$^{-1}$, which is the velocity above which significant ablation of dust by sputtering is believed to occur (Wickramasignhe 1972; Kwok 1975). The lowest expansion velocity for any of the stars in this survey is that of T Ari, 2.7 km s$^{-1}$. While the assumption of uniform spherical grains used to derive equation 4 is a crude approximation at best, this result provides a possible explanation for why no Miras have outflow velocities significantly below 3 km s$^{-1}$.

**Table 4. – Comparison of Outflow Velocity Measurements for Previously Detected Stars**

| | CO Rotational Transition | | | | CO Rotational Transition | | |
|---|---|---|---|---|---|---|---|
| Star | 1–0 | 2–1 | 3–2 | 4–3 | Star | 1–0 | 2–1 | 3–2 | 4–3 |
| T Cas | 10.5 [A] | 5.2 [B] | 10.5 | 10.5 | R Leo | 4.0 [E] | 6.5 [H] | 7.4 | 7.4 |
| W And | 11.0 [C] | ... | 8.9 | ... | R Hya | ... | 7.2 [B] | 10.5 | 8.2 |
| o Cet | 10.0 [D] | 5.0 [E] | 4.8 | 4.7 | W Hya | ... | 8.8 [B] | 8.0 | 8.4 |
| R Hor | 5.4 [D] | 6.6 [F] | 5.2 | ... | R Aql | ... | 10.1 [I] | 9.5 | 9.2 |
| S Pic | 11.8 [D] | ... | 11.6 | ... | $\chi$ Cyg | 10.2 [E] | ... | 9.9 | 9.0 |
| R Aur | 11.1 [D] | ... | 9.6 | 10.0 | T Cep | ... | 5.0 [J] | 4.8 | ... |
| U Ori | 5.8 [G] | ... | 7.5 | 6.5 | R Cas | 14.3 [D] | ... | 12.1 | 13.3 |
| R LMi | 6.5 [D] | 6.0 [H] | 8.7 | 8.5 | | | | | |

Sources: (A) Knapp, 1986; (B) Zuckerman and Dyck, 1986a; (C) Zuckerman, Dyck and Claussen, 1986; (D) Nyman *et al.* , 1992; (E) Knapp and Morris, 1985; (F) Knapp and Sutin, 1989; (G) Kastner, Zuckerman, Dyck and Sopka, 1989; (H) Knapp *et al.* 1982; (I) Wannier and Sahai, 1986; (J) Zuckerman and Dyck, 1989

Although the CO emission lines from these stars presents a radiative transfer problem which is too complex to be solved analytically, it is likely that in all cases the emission from higher J transitions arises from a region closer to the star than does a low J transition. In the case of a large mass loss rate ($\dot{M} \gtrsim 10^{-5} M_\odot \, yr^{-1}$), this is because the rotational levels are populated by collisions (Morris 1980, Schönberg 1988), which tends to force the partition function to reflect a single temperature at a given radial distance. This temperature will decrease with increasing distance from the star, so the Boltzman factor will allow high J levels to be populated only at small radii. For low mass loss rates, collisional excitation is less important than radiative pumping via the CO vibrational transitions near $4.6\mu$ (Morris 1980, Schönberg 1988). Although infrared absorption and subsequent decay can either increase or decrease J, the branching ratios of the transitions insure that an increase in J is more likely, therefore the net effect of infrared pumping is to increase the population of the higher J transitions. However the spontaneous decay rate of a rotational transition is proportional to $J^3$, and as J increases the pump rate must increase also to maintain a significant population. This means that as J increases infrared pumping will only be effective at smaller radii, again implying that high J emission arises from small radial distances. Assuming the CO rotational ladder can be characterised by a temperature distribution of the form $T(r) \propto r^{-\alpha}$, Van der Veen and Olofsson (1989) derived an expression giving the approximate relative locations from which the different rotational emission lines arise

$$\frac{R_{J \to J-1}}{R_{1 \to 0}} = \left( \frac{2}{J(J+1)} \right)^{1/\alpha} .$$

If $\alpha = 0.7$ (Kwan & Linke 1982), this implies the CO(4–3) probes the stellar wind at a radius only $\sim 1/25$ as large as CO(1–0), and reflects the conditions in the wind only a few hundred AU from the star.

Table 4 lists the outflow velocities, culled from the literature, for the stars in this survey which had previously been detected at a lower J transition. One must be cautious in comparing the outflow velocities obtained by different authors who no doubt processed their data with different algorithms. Even for the very strong source Mira, the velocities obtained for a single transition, CO(2–1), show considerable scatter ( Knapp & Morris 1985: $5.0 \, \mathrm{km \, s^{-1}}$; Zuckerman & Dyck 1986a: $4.6 \, \mathrm{km \, s^{-1}}$; Zuckerman, Dyck & Claussen 1986: $8.3 \, \mathrm{km \, s^{-1}}$; Zuckermann & Dyck 1989: $4.3 \, \mathrm{km \, s^{-1}}$). However if the gaseous component of the wind is still being accelerated at radii greater than a few hundred AU, the value obtained for $V_o$ should on average be lower as increasingly high transitions are examined. Table 4 shows no evidence of this.

Two groups have recently studied the acceleration of SiO in oxygen–rich red giants, including $\chi$ Cygni. Lucas *et al.* (1992) mapped SiO(2–1) using the IRAM interferometer, and concluded that the gaseous component of the envelope had attained only 1/2 of its terminal velocity at a distance of 330 AU from the star, and only 3/4 at a distance of 730 AU. In contrast Sahai and Bieging (1993) mapped the same spectral line in several of the same objects using the BIMA array, and concluded that the acceleration zone was only 1/10 as large as Lucas *et al.* had deduced. The results presented here, which show no systematic decrease in the the outflow velocity as higher J transitions of CO are observed, argue in favor of the Sahai and Bieging model.

## 5. Conclusions

A survey of stars from the RCMV within 500 pc was conducted; of the 66 stars examined 36 were detected. The survey data led to the following conclusions:

1) A Mira's spectral type is the best predictor of whether an extensive CO envelope exists or not. None of the stars examined with spectral types earlier than M5.5 were detected, all of those later than M7 were.

2) For the relatively low mass loss rates of these Miras there is no correlation between far infrared color and the mass loss rate indicated by the CO emission, in contrast to the findings of CO surveys containing heavily obscured stars. The far infrared luminosity is, however, correlated to the mass loss rate.

3) Two stars, R Cen and R Car, appear to have extensive dust envelopes, yet were not detected in CO(3–2). Because these stars are near the southern limit of the sky observable from the CSO, it would be useful to examine these stars carefully from an observatory in the southern hemisphere. It would also be interesting to search for atomic carbon emission from these stars, to see if the lack of CO is caused by the stars' relatively warm photospheres, rather than an undetectably low mass loss rate.

4) The outflow velocity increases as stars with cooler photospheres are examined, as would be expected if radiation pressure on dust grains drives the stellar wind.

5) The mass loss rate is strongly correlated to the outflow velocity. A power–law relationship holds between these two quantities over almost 3 orders of magnitude in $\dot{\mathrm{M}}$.

6) At the low end of the observed outflow velocities ($\sim 3 \, \mathrm{km \, s^{-1}}$), the dust drift velocity may be large enough to cause the destruction of the dust grains though dust–gas collisions.

7) At the high end of the observed outflow velocities ($\sim 12 \, \mathrm{km \, s^{-1}}$), the momentum in the wind is beginning to approach the total amount of momentum available in the radiation field. At velocities $\gtrsim 17 \, \mathrm{km \, s^{-1}}$, the power–law $V_o$ – $\dot{\mathrm{M}}$ relationship indicates that all the momentum in the radiation field will be transferred to the wind; this can only occur if the dust envelope is optically thick in the visible and near infrared.

8) An examination of the outflow velocities obtained at differing CO rotational transitions shows no general trend in $V_o$ as high J transitions are observed.

I am greatly indebted to the staff of the CSO for their help and support during the numerous observing runs over which these data were collected. Jill Knapp, Tom Phillips and Antony Schinckel provided helpful suggestions to improve an early version of this paper. Also, I wish to thank the anonymous referee for pointing out several errors I had made. This work was supported by NSF contract #AST 90–15755.